\documentstyle[aps,epsf,preprint]{revtex}
\input {amssym.def}
\input {amssym.tex}
\everymath{\rm}
\everydisplay{\rm}

\begin{document}
\title{Pseudo-Gap Phase, the Quantum-Critical Point \\
and superconductivity in Copper-Oxide Metals}

\author{C. M. Varma}
\address{Bell Laboratories, Lucent Technologies \\
Murray Hill, NJ 07974}

\maketitle

\begin{abstract}
A systematic solution of a general model for Copper-Oxides reveals
a line of transitions $T = T_p (x)$ 
for $x$, the doping away from half-filling
less than a critical value,
to a phase with broken
time reversal and rotational symmetry but with 
translational symmetry
preserved.  The single-particle spectrum in this phase is
calculated to have a gap of $d_{x^2 - y^2}$ symmetry.  The
properties in this phase are compared to the properties of the so
called ''pseudo-gap phase'' in the Copper-Oxides. The fluctuations
towards this phase promote superconductivity which is either of the $d_{x^2-y^2}$ 
or the generalised s-symmetry depending on whether the projected density of
states  at the chemical potential is larger in the $d_{x^2-y^2}$ channel,
as in the hole-doped copper-oxides, or in the s-wave channel. 

\end{abstract}

\newpage
\section*{Introduction}

A schematic phase-diagram of the Copper-Oxide (CuO) 
metals is shown in fig.(1).
The superconducting region is surrounded by three regions 
with distinct properties:  A region marked (III) with 
properties characterestic of a Fermi-liquid, a region marked 
(I) in which a Fermi-surface is discerned but in which the 
quasiparticle concept is inapplicable, and a region marked (II), the
so-called pseudo-gap region, in which the concept of a 
Fermi-surface itself is lost.

The topology of fig.(1) around the superconducting region is that
expected around a quantum critical point\cite{r1} in itinerant Fermions.
The marginal Fermi-liquid (MFL) phenomenology\cite{r2} with which
many of the unusual properties in region (I) are understood assumes
a scale-invariant low energy fluctuation spectrum characterestic of a
quantum-critical point at $x=x_c$, the composition near the highest
$T_c$. 
A region of Fermi-liquid is then expected for $x>x_c$ at low
temperatures, as in region (III), and a region with a broken symmetry
for $x<x_c$ at low temperatures. As in heavy-Fermion compounds,\cite{r3} 
a region of superconductivity is found at low temperatures
peaked in the region around the quantum critical point. From this point
of view, the crucial question in CuO metals is the symmetry of the phase
in the region (II) below $T_p(x)$.

A systematic theory starting with a very general model for CuO 
provides {\it an} answer\cite{r4} to this question.  
The region (II) in Fig. (1) 
is derived to be a phase in which a four-fold pattern of current 
flows in the ground state in each unit-cell as shown in Fig. (2). 
Four-fold symmetry around the copper-atoms as well as 
time-reversal symmetry are broken while their product as 
well as lattice translation symmetry are preserved. 
This phase will be referred to as the Circulating Current (CC) Phase. 
The properties of this phase were not studied in Ref. (4), 
which was mainly concerned with the fluctuations leading to the 
MFL properties in region (I) and the superconductivity induced by them.

In the last few years, the properties in region (II) 
have become much clearer, thanks specially to Angle-Resolved 
Photoemission (ARPES)\cite{r5} and thermodynamic 
measurements.\cite{r6} 
A pseudo-gap in the single particle spectra, consistent with
$d_{x^2-y^2}$ symmetry begins to develop at $T < T_p(x)$ 
which is similar to the temperature at which the specific 
heat coefficient $\gamma (T)$\cite{r6} and the magnetic susceptibility 
$\chi (T)$\cite{r7} begin to decrease and the 
resistivity\cite{r8} and optical\cite{r9} 
and Raman-spectra\cite{r10} 
begin to drop below the MFL behavior of region (I).
I calculate in this paper that the (CC) phase has a single-particle
spectra 
with a gap of    
$d_{x^2-y^2}$ symmetry, as observed, and of the right order of
magnitude, and from which the other properties in region (II) follow. The principal features of the phase-diagram in fig.(1) then arise in
systematic calculations from a single model. I suggest further
experiments to confirm
this identification. (Numerous ideas\cite{gapguys} already suggested for the pseudo-gap phase are discussed elsewhere\cite{batlogg}).

\section*{The Model}

  The CC phase is a mean-field solution of a general Hamiltonian 
in the basis
of three orbitals per unit cell, $d,p_x,p_y$:\cite{r11}
\begin{equation}
H = K + H^{(1)}_{int} + H^{(2)}_{int}
\end{equation}
\begin{equation}
K = \sum_{k, \sigma} \epsilon_d n_{d k \sigma} + 2  \: t_{pd}
d^+_{k, \sigma} \left( s_x (k) p_{x k \sigma} +
s_y (k) p_{y k \sigma}\right) - \: 4 t_{pp} \:
s_x (k) s_y (k) p^+_{x k \sigma} p_{y k \sigma} + H.C.  
\end{equation}
Here a particular choice of the relative phases of 
the x and y orbitals in the unit cell has been made, 
$s_{x,y}(k) = sin (k_x a/2 , k_ya / 2 )$ and for later, 
$c_{x,y}(k)= cos((k_x a/2 , k_y a/2 )$ and 
$s_{xy}^2(k) = (sin^2(k_x a/2 + sin^2 (k_y a/2 ))$.
I consider the local interaction on the Cu and the O orbitals, 
\begin{equation}
H^{(1)}_{int} = \sum_{i, \sigma} U_d \: n_{di \sigma}  \:
n_{di - \sigma} + U_p \left( n_{p x i \sigma} \;
n_{pxi - \sigma} + \; n _{pyi \sigma} n_{pyi - \sigma} \right)
\end{equation}
and the nearest neighbor interaction between the Cu and the O orbitals, 
\begin{equation}
H^{(2)}_{int} = 2V \sum_{k k^{\prime} q, \sigma \sigma^{\prime}}
c_x (q) \: d^+_{k+q \sigma} \: d_{k \sigma}  \:
p^+_{x k^{\prime} - q \sigma^{\prime}} \:
p_{x k^{\prime} \sigma^{\prime}} + x \rightarrow y
\end{equation}
More general interactions do not change the essential 
results derived here. Throughout this paper (renormalised) 
energy difference $\epsilon_d$ between the Cu and the O orbitals is
taken zero. It is important that in CuO, 
$\epsilon_d \lesssim O(t_{pd})$. Taking it as zero simplifies the 
calculations presented; the principal effect of a finite $\epsilon_d$
will be mentioned.

\section*{The Circulating Current Phase}

First, I derive some results of Ref. (4) in a 
simpler way by a simpler treatment of the large on-site 
repulsions in Eq. (3) and calculate the phase diagram of the CC phase.
In the limit
$( U_d,U_p) >> (t_{pd},t_{pp})$, a good mean-field 
approximation\cite{r12} for 
low density of holes (or electrons) consists in replacing 
\begin{eqnarray}
t_{pd} & \rightarrow & \bar{t}_{pd} = t_{pd} |x| \nonumber \\
t_{pp} & \rightarrow & \bar{t}_{pp} = t_{pp} |x| 
\end{eqnarray}
where $|x|$ is the deviation from half-filling in the 
conduction band: $x > 0$ for holes and $x < 0$ for electrons. 
A more general treatment would consider 
separately the average occupation in the oxygen and 
copper orbitals and renormalize $t_{pd},t_{pp}$ accordingly. 
This is an unnecessary complication - no
new physical principle is involved and small quantitative 
modifications are expected since the average occupations 
of Cu and O orbitals in CuO-metals
are quite close (near optimum doping).

A mean-field (non-superconducting) order parameter 
is sought which does not break translational symmetry
(nor the rotational symmetry of spins). 
This means that the mean-field Hamiltonian is just a 
change of the coefficients in the kinetic energy 
operator $K$.  All the possible mean-field decompositions 
of the interactions except one only change the magnitude 
of the coefficients while preserving the symmetry. 
The only interesting mean-field decomposition 
comes from  $H_{int}^{(2)}$ and yields the complex 
mean-field order parameter
\begin{equation}
Re^{i \phi} \equiv V/2\sum_{k \sigma}
s_x (k) \left< d^+_{k \sigma} \: p_{x k \sigma} \right> - s_y (k)
\left< d^+_{k \sigma} \: p_{y k \sigma} \right>
\end{equation}
The mean-field Hamiltonian itself is
\begin{equation}
H_{mf} = K - Re^{-i \phi} \sum_{k \sigma} \left(s_x(k) \:
d^+_{k \sigma} \: p_{x k \sigma} - s_y (k) \:
d^+_{k \sigma} \: p_{y k \sigma}\right) + H.C. + 
\frac{R^2}{2V}
\end{equation}
By a unitary transformation 
\begin{equation}
d_{k \sigma} \rightarrow d_{k \sigma} , \: p_{xk \sigma}
\rightarrow p_{xk \sigma} \: e^{i \theta_x }, 
p_{yk \sigma} \rightarrow p_{y k \sigma} \:
e^{i \theta_y },
\end{equation}
\begin{equation}
tan \theta_{x,y} = \pm Rsin \; \phi \; /(2 \bar{t}_{pd} \pm Rcos \; \phi )
\:, 4 \hat{t}^2_{pdx,y} = \left( 2 \bar{t}_{pd} \pm R cos \phi \right)^2
\; + R^2 sin^2 \phi, 
\end{equation}
the phase difference is transferred completely to 
the $p_x - p_y$ bond, so that $H_{mf} \rightarrow H_{\theta}$,
\begin{equation}
H_\theta = \sum_{k \sigma} 2\left[ 
d^+_{k \sigma} \left( \hat{t}_{pdx}s_x (k) p_{x k \sigma} +
\hat{t}_{pdy}
s_y (k) p_{y k \sigma} \right) \right] - 4 \bar{t}_{pp} \:
e^{i \theta} \nonumber \\ 
s_x(k) s_y(k) \: p^+_{x k \sigma} \:
p_{y k \sigma} + H.C. + \: R^2/2V \:,
\end{equation}
where $\theta = \theta_x - \theta_y$.
It is apparent that only the phase difference 
of the two $p-d$ bonds in the unit cell is gauge-invariant. 
With the most general interaction terms, the gauge-invariant 
phase is the change in phase going around the plaquette formed 
by a Cu-atom and any two of its oxygen neighbors, which differs 
from the above only in the value of $\theta $. 
 
The energy of each of the three bands obtained by diagonalising Eq. (7) is changed by a 
finite $(R,\phi )$.  But for any $k$ the trace of the change 
in energy of the three bands is zero.  The change in energy 
can therefore be expressed purely in terms of the change 
in energy $\delta \: \epsilon_{ck}$ of the filled part of 
the (hole) conduction band. So the mean-field values $R_0$ and
$\phi_0$   are determined 
by minimizing
\begin{equation}
\frac{R^2}{2V} - 2 
\sum_{k} \delta \epsilon_{ck} 
\: ( \phi , R ) \:
f ( \epsilon_{ck} ).
\end{equation}
 $\delta \epsilon_{ck}$ has a complicated expression. To $O(R^2)$ and leading order in $(\bar{t}_{pp} / \bar{t}_{pd})$ it is   
\begin{equation}
\delta \epsilon_{ck} \approx \left(\frac{R}{2\bar{t}_{pd}}\right)^2 \left(  \bar{t}_{pd} s_{xy}^2 (1-1/4 cos^2\phi) +
\frac{4(\bar{t}_{pp} s^2_x (k) s^2_y (k)) }{s^2_{xy (k)}}
 \:  ( 1 - \frac{1}{2} \cos^2 \phi) \right).
\end{equation}

From Eq. (11) and (12) $R_0 \neq 0$ at $T=0$ for 
\begin{equation}
\frac{2|x|\hat{t}_{pd}}{V} <  \sum_{k < k_F} \left(s_{xy}^2(k) +
\frac{8 t_{pp}}{t_{pd}}
\frac{s^2_x (k) s^2_y (k)}{s^2_{xy} (k_)}\right),
\end{equation}
and  $\phi_0$ is $\pi/2$ or $-\pi/2$. The symmetry of the transition is therefore of the Ising variety. 

For $|x|<<1$, Eq.(13) is satisfied
only for $x$ less than critical doping $|x_c|$, 
\begin{equation}
|x_c| \approx 
\frac{1}{2} (V /(t_{pd})(0.25 + 0.5 t_{pp} 
/ t_{pd}).
\end{equation}
$|x_c|$ defines the 
quantum critical points for bothe electron and hole dopings.
An estimate of $\theta_0 \equiv R_0/(2\bar{t}_{pd})$  obtained by expanding Eq. (12) to $O(R^4)$ is
\begin{equation}
\theta_0^2 = O\left( \bar{t}_{pd}^2/2[V (t_{pp}+t_{pd})])(x_c-x)\right ).
\end{equation}
For $\epsilon_d \neq 0$, $t^2_{pd}$ is replaced by 
$t^2_{pd} + O ( \epsilon^2_d )$. With $V, t_{pp} ,t_{pd}$, and $\epsilon_d$ 
of similar magnitude, as for CuO compounds, $x_{ec},x_{hc}$ 
are therefore about 0.2. The line of transitions at finite 
temperatures varies with $x$ as 
\begin{equation}
T_p /E_F \sim |x_c - x|^{1/2}   
\end{equation}
This
is to be identified with $T_p(x)$ of fig.(1). The competition with 
antiferromagnetism near $x=0$ is not treated here.

The energy of the 
(transverse) fluctuations in $\phi$ about $\phi_0$ at 
long wavelengths is estimated from Eq. (12) to be
\begin{equation}
\Omega_0^0 = O \left( \theta_0^2 (\bar{t}_{pp} +\bar{t}_{pd})/4\right ).
\end{equation}
The variation of $\Omega_q^0$ with q is estimated to be slow, of $O(qa)^2$.

The eigenvectors of the states in the conduction band to 
leading order in $\theta$ and $t_{pp}/t_{pd}$ are
\begin{equation}
| c_{k \theta \sigma} > \: = \:
\frac{1}{N_k} \left[ | c_{k o \sigma} >
+ i2\sqrt{2} \theta_0\frac{\bar{t}_{pp}}{\bar{t}_{pd}} \;
s_x (k) s_y (k)  
| a_{k o \sigma} > 
 \right]
\end{equation}
where $| a_{ko} >$ and $| c_{ko} >$ are the
nonbonding and conduction band states of $k$ for
$\theta = 0$ and $N_k$ is the normalization factor.
The term proportional to 
$\theta_0$ leads to a difference in the phase between the
$| p_{x k \sigma} >$ and $| p_{y k \sigma >}$ 
terms in Eq. (18) producing
a time-reversal breaking state.
Similarly expressions can be derived for the other two bands. 
The three together correspond to a current carrying state 
with pattern shown in Fig. (2). The wave function of states with 
$\phi_0 =-\pi/2$ is (18) with $-i\theta_0$ in the second term. Note that
states with $\phi_0 = \pi/2$ and $-\pi/2$ are not orthogonal. But the         two ground 
states formed from the Hartree-Fock product of such states are orthogonal.

The CC phase breaks a rotational invariance of the Cu-O lattice
(besides the time-reversal invariance).  So lattice defects, such
as interstitials, dislocations or grain boundaries couple as
external fields do to a $d = 2$ Ising order 
parameter.\cite{r13}  No
thermodynamically sharp transition is therefore possible at 
$T_p(x)$.

\section*{Anisotropic Gap in the Normal state}

In the mean-field approximation above, the energy of states in the 
conduction band shift in a k-dependent fashion, Eq. (12) 
with nothing special happening at the chemical potential. This is because the
source of the instability considered is interband transitions. 
This solution will now be shown to be modified 
by the fluctuations which lead to scattering among the conduction band 
states, (18), near the chemical potential. 
The fluctuations to be considered 
are transverse, i.e. of $\phi$. 
whose energy $\Omega_0$ at long wave-lengths is 
given by Eq.(17). The Hamiltonian for such fluctuations, 
generated by operators $\delta \phi_q , \delta \phi_q^+$,
 is quite generally obtained 
from the deviation of the starting 
Hamiltonian from $H_{mf}$.
\begin{equation}
H_{fluct} = \sum_q \Omega_q^0 \; \delta \phi^+_q 
\; \delta \phi_q 
+ \sum_{kk^{\prime} \sigma} g (k,k^{\prime}) \: c^+_{k^{\prime} \theta
\sigma} 
\: c_{k \theta \sigma}
\left( \delta \phi_{k-k^{\prime}} + \delta \phi^+_{k^{\prime}-k} \right)
+ H.C.
\end{equation}
The coupling function to zeroeth order in $t_{pp}/t_{pd}$ is calculated to be
\begin{equation}
g(k,k^{\prime}) = \theta_0 \bar{t}_{pd}
[s_x(k)s_x(k^{\prime})-s_y(k)s_y(k^{\prime})]/s_{xy}(k).
\end{equation}
For $k \rightarrow k^{\prime}$,this is $\sim (cos(k_xa)-cos(k_ya))$.

Consider now the renormalisation of the phase-fluctuation energy through
Eq.(19). For small q
\begin{equation}
\Omega^2(q) = \Omega_q^{0 2} - 2\Omega_q^0 \bar{t}_{pd}^2\theta_0^2
(1-q^2/8) \sum^{\prime}\left( (cos(k_xa)-cos(k_ya)/s_{xy}\right)^2
/ ( \epsilon_{k+q} - \epsilon_k ),
\end{equation}
with the restriction on the sum that $k<k_f$ and $k+q>k_f$. The self-energy of the fermions at $k=k_f$ and energy $\omega$ is
$\sim ln|\omega-\Omega_0^0|$ suggesting an instability if the
fluctuation energy at long wavelength approaches 0. The conclusion from Eq.(21) and Eq. (17) is that the phase fluctuations can be   
unstable over a substantial part of momentum-space. A way to 
restore stability is by a condensation of the phase 
fluctuations and a corresponding modification of the electronic energy.   

The condition for stability is derived by the ansatz that $<\delta \phi_q>$
and $<c_{k+q,\theta \sigma}^+c_{k\theta \sigma}>$ are finite for a range of
$q$ around zero \cite{polar}. Eq. (19) then leads to new eigenstates which are     annihilated/created by $\gamma_{\it{k}\sigma}$, $\gamma_{\it{k}\sigma}^+$: 
\begin{equation}
\gamma_{\it{k}\sigma} = (c_{k\theta \sigma} + \sum_{q\ne 0}u_{{\em k},k+q} c_{k+q\theta \sigma})/M_{\it{k}},
\end{equation}
where $M_{\it{k}}$ is the normalisation. It will turn out that
$u_{\it{k},k+q} \ne 0$ only for states with $|(\epsilon(k),\epsilon(k+q))-\mu | \lesssim \Omega_0^0$ . For such states momentum is not a good quantum number; I label the new states by {\em k} to indicate that the {\em average} momentum of such a state is $k$. In this sense, the problem retains translational invariance {\em only on the average}.

From Eq. (19),
\begin{equation}
<\delta \phi_q> = -\frac{1}{\Omega_q^0} \sum_k^{\prime} g(k,k+q) <c_{k+q,\theta \sigma}^+c_{k\theta \sigma}>.
\end{equation}
I use the the Brillouin-Wigner (B-W) self-consistent approximation to get:
\begin{equation}
<c_{k+q,\theta \sigma}^+c_{k\theta \sigma}> = g(k,k+q)<\delta \phi_q>/
(E_{\it{k}}-E_{\it{k+q}}),
\end{equation}
where $E_{\it{k}}'s$ are the new one-particle eigenvalues to be determined. The B-W approximation is exact in the limit that the number of states $(k+q)$ coupled to a given state $k$ is very large compared to 1, as here.

Combining Eqs.(23) and (24) yields the self-consistency equations 
\begin{equation}
1 = \Omega_q^{0-1} \sum_k^{\prime} |g(k,k+q)|^2/(E_{\it{k+q}}-E_{\it{k}}),
\end{equation}
\begin{equation}
\Phi({\em k,k+q}) = g(k,k+q)/\Omega_q^0 
\sum_{k^{\prime} \sigma^{\prime}}^{\prime}
g(k^{\prime},k^{\prime}-q)/(E_{{\em k^{\prime}-q}}-
E_{{\em k}}^{\prime})\Phi ({\em k^{\prime},k^{\prime}-q}), 
\end{equation}
where
\begin{equation}
\Phi({\em k,k+q}) = (E_{{\em k+q}}-E_{{\em k}})<c_{k+q \theta \sigma}^+c_{k \theta \sigma}>.
\end{equation}
In Eqs. (23)-(26), the sum is restricted to states k etc. such that $\epsilon(k)$ is within about $\Omega_0^0$ of the chemical $\mu$, as required by the retarded nature of the Fermion-Boson interaction. 
The restriction on the sums is more clearly seen if Eq.(19) 
is used to generate an
effective frequency dependent Fermion vertex
\begin{equation} 
\sum_{k,k^{\prime},q,\sigma,\sigma^{\prime}}
g(k,k+q)g(k^{\prime}-q,k^{\prime})D_{\phi}(q,\omega)
c_{k+q \theta \sigma}^+c_{k \theta \sigma}
c_{k^{\prime}-q \theta \sigma^{\prime}}^+
c_{k^{\prime} \theta \sigma^{\prime}}.
\end{equation}
where $D_{\phi}(q,\omega)$ is the propagator for the fluctuations which on the energy shell may be
approximated by $-1/\Omega_q^0$
with a cut-off such that
$\epsilon(k^{\prime}),\epsilon(k^{\prime}+q)$
are both within about $\Omega_0^0$ of $\mu$. 
A quadratic
hamiltonian obtained by taking the expectation value of
$<c_{k+q,\theta \sigma}^+c_{k\theta \sigma}>$ etc. yields (26) using the B-W
approximation.

Eqs.(26),(27) have been obtained from the 
"direct" contraction of Eq (22). The exchange contraction yields
a similar equation with a coefficient which, in the relevant channel, is $(-1/4)$ of that on the right side of Eq (26). 
 
 For an (approximate) solution of Eqs.(25),(26), note first that since
$q \rightarrow 0$, $g(k,k+q) \sim (cos(k_xa)-cos(k_ya))$, for $q \rightarrow 0$, $\Phi(k,k+q) \sim (cos(k_xa)-cos(k_ya))$ also. Second, $E(k) \approx \epsilon(k)$ for $|\epsilon(k) - \mu | \gtrsim \Omega_0^0$. 
A consistent 
solution for k near $k_f$ is then
\begin{eqnarray}
E_{{\em k}} &=& \epsilon_k + D(\it{k}), k \gtrsim k_f  \nonumber \\
E_{{\em k}} &=& \epsilon_k - D(\it{k}), k \lesssim k_f  \nonumber \\
D(\it{k}) &=& D_0(cos(k_xa)-cos(k_ya))^2,
\end{eqnarray}
as may be verified by substitution. 
The value of $D_0$ is evaluated to be of
$0(\theta_0^2 \; \bar{t}_{pd}^2 \rho (0))$. 

The gap in the one particle spectrum at the 
chemical potential $D(k)$ has a $d_{x^2-y^2}$ symmetry and 
a magnitude which can be estimated from 
Eqs. (15) to be $x(x_c-x)t_{pd}^2(t_{pd}+t_{pp})/(Vt_{pp})$
 which at say
$(x_c - x ) \approx .05$, and $t_{pd} \sim V \sim 
t_{pp} \sim 1eV$  is 
$\approx 20$ meV.  
This has the right order of magnitude\cite{r5}. The differences of (29) from 
D-wave BCS single-particle spectrum are significant and have observable
consequences discussed below.

The leading decrease in energy due to the modification of the
single-particle spectrum is $\sim \theta_0^2$.  Therefore, the
nature of the transition remains unchanged,at least in mean-field theory; only quantitative
changes are introduced in the conditions Eqs. (16) and (17)
for the occurence
of the CC phase.

\section*{Properties in the Circulating Current Phase}

The single particle density of states in the CC phase calculated 
 from Eq. (29) is
\begin{eqnarray}
\rho_{cc} ( \omega ) &=& \rho (0) \frac{2}{\pi} Arcsin 
\left( \left| \frac{\omega}{D} \right|^{1/2} \right) ,
\left| \frac{\omega}{D} \right| \leq 1 \nonumber \\
&=& \rho (0) , \left| \frac{\omega}{D} \right| \geq 1
\end{eqnarray}
This increases as $\left| \omega / D \right|^{1/2}$ for 
$\left| \omega / D \right| << 1$ and unlike the d-wave
superconductors, (which have a logarithmic singularity at
$\omega = \Delta$, the superconducting gap), 
$\rho_{cc} ( \omega )$ is less than the "normal" density of states
$\rho(0)$ at all $| \omega | < D$.

 Eq. (30) should be compared with the single particle density of
states measured in tunneling\cite{r14} 
and with the specific heat\cite{r6} $C_v$ and
magnetic susceptibility\cite{r7}  $\chi$.  The former shows, (just as Eq
(30)),
a diminuition in the single particle density of states for low
energies at $T \lesssim T_p (x)$ but show no rise 
above $\rho (0)$ at finite
energies until $T \lesssim T_c (x)$, when the characteristic
superconducting density of states appears.  $C_v$ in the CC phase
is predicted $\sim T^{3/2}$ and $\chi \sim T^{1/2}$ for
$T << T_p (x)$.  Due to the intervention of superconductivity, it
is hard to test these power laws accurately.  In the measured
range\cite{r5,r6} $T \chi / C_v$ is 
nearly independent of temperature as
predicted here.  $\chi (T)$, measured more accurately than
$C_v (T)$, can be fit to $T^{1/2}$.  One can deduce the
continuation to the T-dependence below $T_c$ by invoking
conservation of entropy on the $C_v (T)$ measurements.
$C_v (T) \sim T^{3/2}$ for $T << T_p (x)$ is then not inconsistent
while $C_v (T) \sim T$ clearly is.

To obtain the spectral function $A (k, \omega )$
measured by ARPES\cite{r5} 
one needs besides Eq. (29), the 
self-energy of the single particle states (and the `coherence factors').
The situation is in some ways similar to but in an important way
different from the corresponding calculation for a d-wave
superconductor\cite{r15} (and completely different from s-wave
superconductors).\cite{r11}  In both cases, the bare polarizability
$\chi_0 (q, \omega )$ (calculated from a single particle-hole
bubble) is zero for $\omega < D (q)$, ($\Delta (q)$ for the
superconductors).  The lowest energy single particle scattering
for momentum $q$ occurs by an intermediate one-particle state near
the zero of the gap.  Therefore the threshold for single-particle
scattering of a state at $q$ is also $D (q)$.  For
$\omega \gtrsim D (q)$, $\chi_0 (q)$ with 
$E_k$'s given by Eq. (29) is proportional to
$\frac{\omega}{D\epsilon_F}$. So the renormalized
$\chi (q, \omega )$ for $\omega > D (q)$ is expected to be 
similar to the
normal state above $T_p (x)$, i.e. of the marginal
Fermi-liquid form.  
The single particle self-energy needs to be calculated in detail
but its general form is evident:  For 
$ ( \omega , T ) << D (q)$ it is exponentially small, but for
$( \omega , T ) \gtrsim D (q)$
it returns to the value $\Sigma_n( \omega , q , T)$ without the
pseudo-gap.  So
\begin{equation}
Im \Sigma ( \omega , q, T) \approx \; \mbox{sech} \;
\left( \frac{D (q,T)}{( \omega^2 + \pi^2 T^2 )^{1/2}}\right)
Im \Sigma_n ( \omega , q, T) \;.
\end{equation}
The spectral function at the Fermi-wave-vectors $\hat{k}_f$ defined by
$E(\hat{k}_f)= \pm D(\hat{k}_f)$ calculated using Eq. (29) and (31) and
the marginal Fermi-liquid form for $\Sigma_n$ is plotted in fig.(3) for
a few temperatures \cite{projection}.  A pseudogap in the direction $\hat{k}_f$ appears
below $T \approx D(\hat{k}_f)$ producing the illusion of `Fermi-arcs'
shrinking as temperature decreases. The lineshape in fig. (3) at low
energies,
in the pseudo-gap region, fits the experimental curves \cite{r5} 
well within the experimental
resolution. A prediction following from the results of the previous section
is that $D \sim (x_c-x)$ and also $ \sim (T_p(x) - T)$.

The single particle spectrum in $d = 2$ d-wave
superconductor\cite{r15} only differs from the above because of a
singularity in $\chi_0 (q, \omega )$ for $\omega \approx 2
\Delta$ and $q\approx 2k_f$ in the directions connecting the maximums of the gap. This leads to a pole in $Re \Sigma ( \omega , q )$ and
therefore a sharp peak in 
$A ( \omega , q )$ at $q$ on the Fermi-surface close to the maximum of the gap followed by a continuum starting at $\Delta $. No such sharp feature is to be expected in the pseudo-gap region because no singularity in               $\chi_0 (q, \omega )$ exists.

\section*{Superconductivity}

 I discuss here the superconducting instability in the quantum fluctuation
regime (Region I in Fig. (1)) in which $\theta_0 = 0$ and effective electron-electron interaction is through fluctuation in  $\theta$. (The calculation in the CC phase is
similar with the added complication that one must do pairing
calculations in a gapped electronic structure.) The coupling of the
$\theta$ fluctuations
to the Fermions has the same functional form as $g(k,k^{\prime})$ and
leads
to an effective pairing interaction in the spin-singlet channel:
\begin{equation}
H_{pair}= \sum_{k,k^{\prime}} |V(k,k^{\prime}|^2
D_{\theta}(k-k^{\prime},
\omega ) c^+_{k^{\prime},\sigma}
c^+_{-k^{\prime},-\sigma}c_{-k,-\sigma}c_{k,\sigma}
\end{equation}
Here $D_{\theta}$ is the propagator of the $\theta$ fluctuations,
which is essentially momentum independent and has a 
frequency dependence of a scale-invariant form\cite{r4}. We may approximate it
below its
cut-off frequency by $-(\omega_c)^{-1}$ The coupling
function in Eq (32) is then expanded in a separable form in
lattice-harmonics as follows: 
\begin{equation}
V_{pair}(k,k^{\prime}) = -V^2/(4\omega_c) \sum_i g_i
\eta_i(k) \eta_i(k^{\prime})
\end{equation}
where i are the generalised $s$, $d_{x^2-y^2}$, and $d_{xy}$ channels:
\begin{eqnarray}
\eta_s(k) &=& s_{xy}^2(k), g_s=1/2  \nonumber  \\
\eta_{x^2-y^2}(k) &=& 1/2[cos(k_xa)-cos(k_ya)], 
g_{x^2-y^2} = 1/2 \nonumber \\
\eta_{xy}(k) &=&  s_x(k)s_y(k), g_{xy} = -2.
\end{eqnarray}

The methods of Ref.(3), may be used now to deduce that
pairing is possible only generalized s-wave and the $d_{x^2-y^2}$
channels. The choice between the channels is determined by the
band-structure 
near the Fermi-surface. In the hole-doped cuprates $d_{x^2-y^2}$ is
favored because in this symmetry the gap occurs over regions of
Fermi-surface
where the bands-disperse the least (because of the proximity to the Van
Hove
critical point) thereby reducing the energy optimally. If $\mu$ is far away from the van Hove singularity, as in electron-doped cuprates\cite{anlage} the projected
generalised s-wave density of states is expected to be larger than the
projected $d_{x^2-y^2}$ density of states and s-wave pairing is favored.

\section*{Concluding remarks}

Although, I have presented a systematic theory in
agreement with the principal experimental results, one can be
confident of the applicability of the theory only if the structure in
Fig. (2) is observed.  I have suggested\cite{r4} 
elastic polarized neutron
scattering to observe the orbitals moments in Fig. (2).  Another
way to test Fig. (2) is through the polarization dependence of
ARPES in single-domain samples. Some other predictions are mentioned in the text.

Improvements could be made in the theory, especially the derivation of the gap
in the Circulating-current phase. Note that the conditions (26,27) are equivalent to requiring that the phase-fluctuation modes have zero-energy as $q\rightarrow
 0$, which appears to give an $xy$-symmetry character to the circulating current state. If true, no specific heat anomaly is found even in the pure limit
(except for three-dimensional couplings). The ansatz, Eq.(22), by which the 
problem has been diagonalised may be looked upon as admixing states with 
$\phi_0 = -\pi/2$ in to states with $\phi_0 = \pi/2$. It is possible that a better treatment would express the one-paticle states as topological excitation  of fermions bound to co-moving current fluctuation. These matters require further investiagation.

I wish to thank E. Abrahams, B. Batlogg, P. Majumdar, and A. Sengupta for useful
discussions.

\newpage
\section*{Figure Captions}
\begin{itemize}
\item[Fig. (1):]
Generic phase-diagram of the cuprates for hole-doping.  Not shown
is a low temperature ''insulating phase'' in region II due to
disorder.
\item[Fig. (2):]
Current pattern predicted in phase II of Fig. (1).
\item[Fig. (3):]
The spectral function at the Fermi-wave-vectors as a function of energy
below the chemical potential normalized to the gap in the direction of
that wave-vector $D(\hat{k}_f)$ for temperatures
$T=nD(\hat{k}_f)/(4\pi)$ for
$n=1,2,3,4,5$.
\end{itemize}

\begin{references}
\bibitem{r1}
See, for example, S. Sachdev, {\it Quantum Phase Transitions},
Cambridge University Press (1999).
\bibitem{r2}
C. M. Varma, et al., Phys. Rev. Lett. {\bf 63}, 1996 (1989), C. M.
Varma, Int. J. Mod. Phys. {\bf 3}, 2083 (1989).
\bibitem{r3}
N. Mathur, et al., Nature, {\bf 394}, 39 (1998).  The proximity of
superconductivity to antiferromagnetism in the heavy Fermions was
noted earlier by K. Miyake, S. Schmitt-Rink and C.M. Varma, Phys. Rev. B {\bf 34}, 6554
(1986), where a d-wave superconductivity from antiferromagnetic
fluctuations was derived.  
\bibitem{r4}
C. M. Varma, Phys. Rev. {\bf55}, 14554 (1997).
\bibitem{r5}
D. S. Marshall, et al., Phys. Rev. Lett. {\bf 76}, 4841 (1996); H.
Ding, et al., Nature {\bf 382}, 51 (1996); A. G. Loeser et al.,
Science {\bf 273}, 325 (1996); M. R. Norman et al., Nature 
{\bf 392}157 (1998); A.V. Federov et al.,Phys. Rev. Lett.{\bf 82},
 2179 (1999). 
\bibitem{r6}
J. W. Loram, et al., Phys. Rev. Lett. {\bf 71}, 1740 (1993).
\bibitem{r7}
M. Takigawa, et al., Phys. Rev. B {\bf 43}, 247 (1991);
H. Alloul, et al., Phys. Rev. Lett. {\bf 70}, 1171 (1993);
J. W. Loram, et al., {\it Proceedings 10th Anniversary HTS
Workshop}, (World Scientific, Singapore (1996)).
\bibitem{r8}
See for example, B. Batlogg, et al., Physica C {\bf 235}, 130
(1994).
\bibitem{r9}
A. V. Puchkov, et al., Phys. Rev. Lett. {\bf 77}, 3212 (1996).
\bibitem{r10}
R. Nemtschek, et al., Phys. Rev. Lett. {\bf 78}, 4837 (1997);
G. Blumberg, et al., Science {\bf 278}, 1427 (1997).
\bibitem{gapguys}
M.R. Norman et al., Phys. Rev. {\bf B57}, R11093, (1998);
V.J. Emery and S.A. Kivelson, Nature {\bf 374}, 434 (1995):
X.G.Wen and P.A. Lee, Phys. Rev. lett. {\bf 80}, 2193 (1998); 
C. Castellani et al., Z. Phys B{\bf 103}, 137 (1997); L.Balents, M.P.A.Fisher,
 and C.Nayak, Int.J. Mod. Phys.{\bf B12},1033 (1998).
\bibitem{batlogg}
B. Batlogg and C.M. Varma (preprint).
\bibitem{r11}
C. M. Varma, S. Schmitt-Rink, and E. Abrahams, Solid State Comm.,
{\bf 62}, 681 (1987).
\bibitem{r12}
The effects of large repulsion can be expressed as a modification
of the kinetic energy by a number of different methods - the most
popular is the slave Boson method, see for example, N. Read and D.
M. Newns, J. Phys. C {\bf 16}, 3273 (1983).

\bibitem{r13}
Y. Imry and S. K. Ma, Phys. Rev. Lett. {\bf 80}, 149 (1998);
\bibitem{polar}
I have taken the limit $\Omega << v_fq$ in evaluating the correction to the frequency by coupling to the Fermions. In the opposite limit the correction is proportional to $q^2$. So, (21) may be considered valid only for $q \gtrsim
\Omega /v_f$ and the indicated instability is maximal at finite $q$. This $q$
however is so small that curing the instability without breaking translational invariance {\em on the average} is reasonable.
\bibitem{r14}
Ch. Renner, et al., Phys. Rev. Lett. {\bf 80}, 149 (1998);
N. Miyakawa, et al., Phys. Rev. Lett. {\bf 80}, 157 (1998).
\bibitem{r15}
H. Y. Kee and C. M. Varma, Phys. Rev. B {\bf 58}, 15035 (1998).
\bibitem{r16}
P. B. Littlewood and C. M. Varma, Phys. Rev. B {\bf 46}, 405
(1992).
\bibitem{projection}
This calculation has ignored the correction arising from the `coherence factor'
in projecting the free-particle states to states created by $\gamma_{\em k}^+$
etc. which is an additional source of broadening.
\bibitem{anlage}
S.M. Anlage et al., Phys. Rev. B {\bf50}, 1994 (1994).
\end{references}
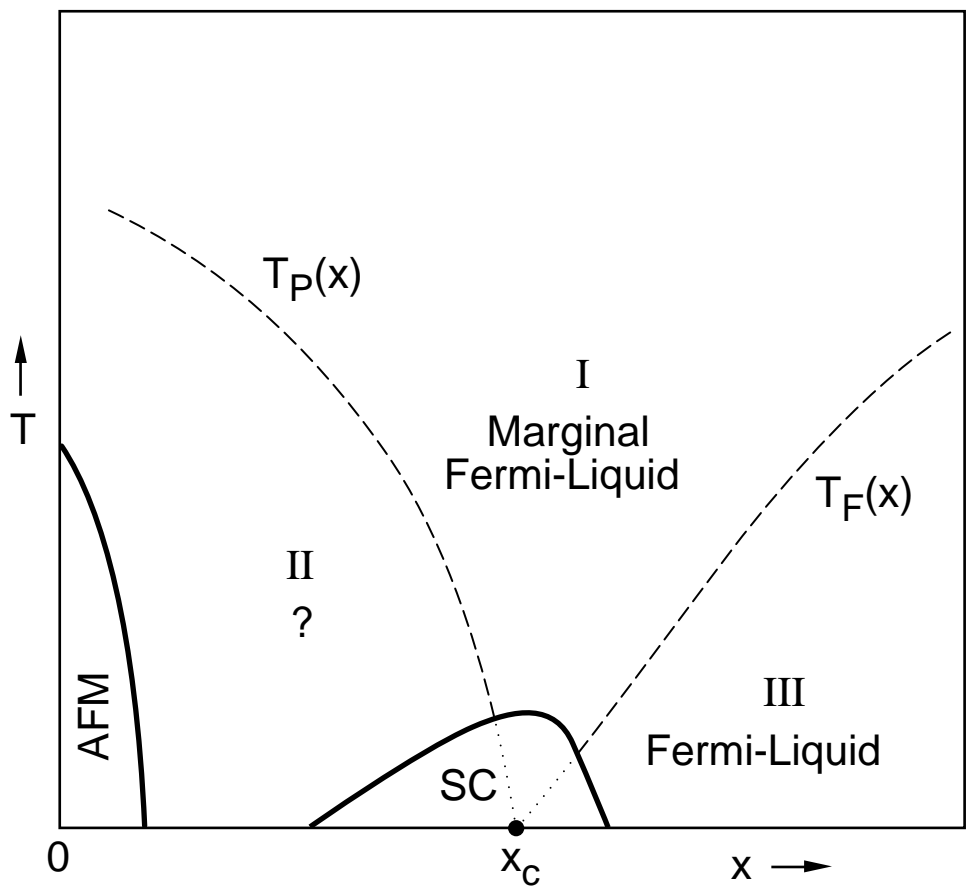
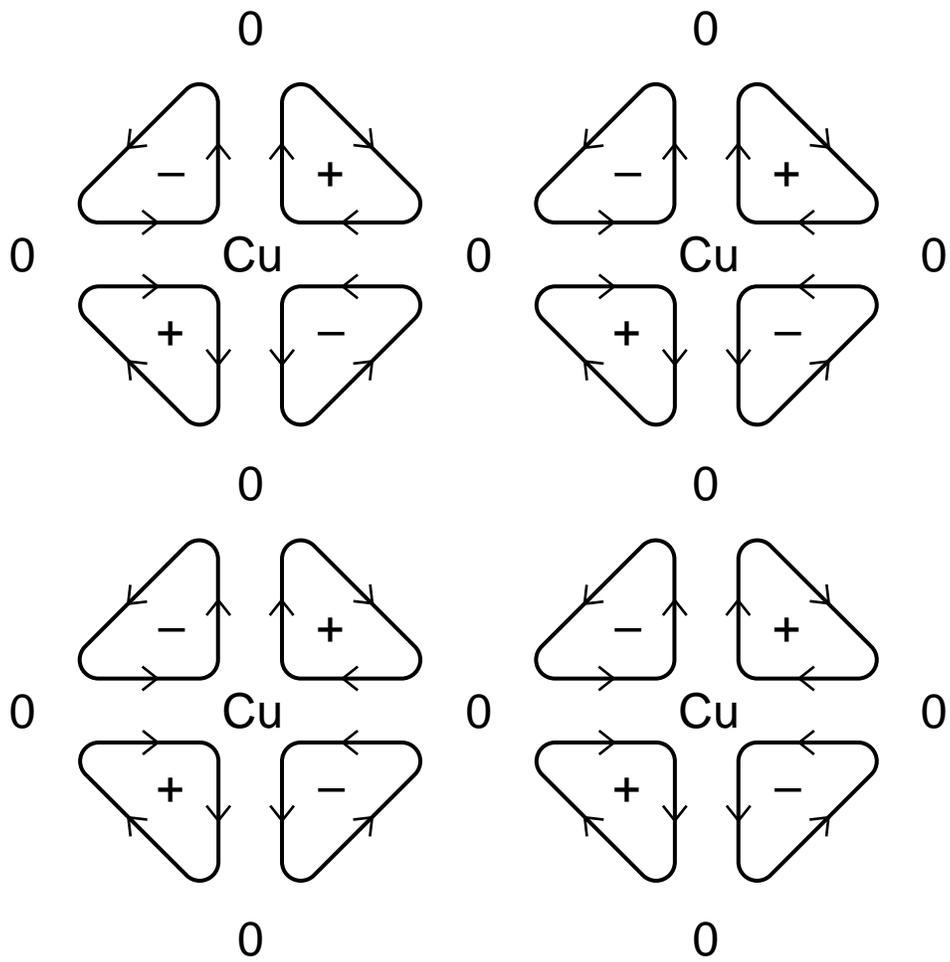
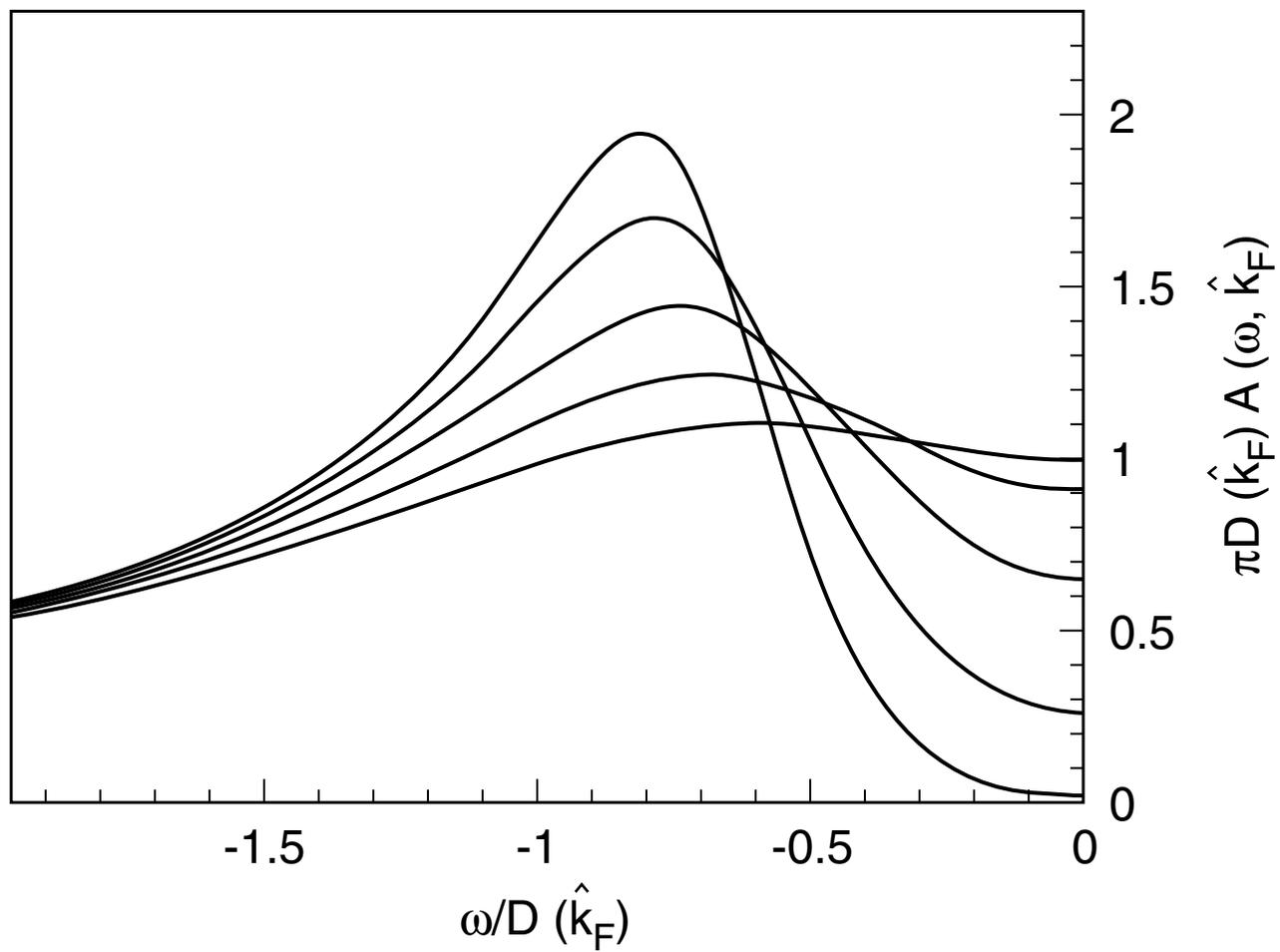
\end{document}